\documentclass[preprint,prd,noshowpacs,nofootinbib]{revtex4}

\usepackage{color} 
\usepackage{amssymb}
\usepackage{amsmath}
\usepackage{graphicx} 
\usepackage{float}
\usepackage{braket}     
\usepackage{pdfpages}
\usepackage{epstopdf}
\usepackage[utf8]{inputenc}

\begin{document}

\title{``The more things change the more they stay the same''\\ {\small Minimum lengths with unmodified uncertainty principle and dispersion relation}\footnote{Essay written for the Gravity Research Foundation 2022 Awards for Essays on Gravitation. \\
2$^{nd}$ Place Award \\}}

\author{Michael Bishop}
\email{mibishop@mail.fresnostate.edu}
\affiliation{Mathematics Department, California State University Fresno, Fresno, CA 93740}

\author{Joey Contreras}
\email{mkfetch@mail.fresnostate.edu}
\affiliation{Physics Department, California State University Fresno, Fresno, CA 93740}

\author{Douglas Singleton (Corresponding author)}
\email{dougs@mail.fresnostate.edu}
\affiliation{Physics Department, California State University Fresno, Fresno, CA 93740}

\date{\today}

\begin{abstract}
Broad arguments indicate that quantum gravity should have a minimal length scale. In this essay we construct a minimum length model by generalizing the time-position and energy-momentum operators while keeping much of the structure of quantum mechanics and relativity intact: the standard position-momentum commutator, the special relativistic time-position, and energy-momentum relationships all remain the same. Since the time-position and energy-momentum relationships for the modified operators remains the same, we retain a form of Lorentz symmetry. This avoids the constraints on these theories coming from lack of photon dispersion while holding the potential to address the Greisen-Zatsepin-Kuzmin (GZK) puzzle of ultra high energy cosmic rays.          
\end{abstract}

\maketitle

{\large {\bf Quantum gravity and a minimal length resolution}} \\

Different research threads lead to the idea that combining quantum mechanics with gravity will give rise to a minimum distance scale or a minimum spatial resolution  \cite{vene,amati,amati2,gross,maggiore,garay,KMM,adler,scardigli}. From the quantum mechanical side, Heisenberg's uncertainty principle implies an inverse relationship between spatial resolution and momentum resolution:  $\Delta x \sim \frac{\hbar}{\Delta p}$. This gives no minimum in the spatial resolution  since $\Delta x$ can be made arbitrarily small by making $\Delta p$ arbitrarily large. Once gravity enters the picture, something must change. To resolve smaller $\Delta x$ means packing more mass-energy into a smaller and smaller spatial region. At some point enough mass-energy will be so densely packed that a black hole will form with a spatial resolution set by the Schwarzschild radius $\Delta x \sim r_{Sch} \sim  G M \sim  G \Delta p $. \footnote{We set $c=1$ throughout this essay which mean mass, energy, and momentum are interchangeable.}  Combining these two ideas leads to $\Delta x \sim \frac{\hbar}{\Delta p} + G \Delta p$ which has a minimum in $\Delta x$ at $\Delta p_m \sim \sqrt {\hbar /G}$ given by $\Delta x_m \sim \sqrt{\hbar G}$. The relationship between $\Delta x$ and $\Delta p$ is shown graphically in figure 1. 
\begin{figure}[H]
\label{fig1}
    \centering
    \includegraphics[scale=0.80]{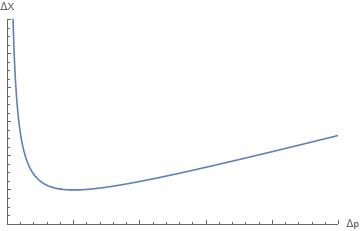}
    \caption{The graph of the generalized uncertainty given by $\Delta x \sim \frac{\hbar}{\Delta p} + G \Delta p$. After the minimum $\Delta x$ is reached there is a linear increase with $\Delta p$.}
\end{figure}

Working from this heuristic argument for a minimum spatial resolution, one well-studied approach \cite{KMM} is to modify the Heisenberg uncertainty principle so as to yield the relationship between $\Delta x$ and $\Delta p$ given above. The generalized uncertainty principle (GUP) in \cite{KMM} has an inequality of the form $\Delta x \Delta p \geq \frac{\hbar}{2}(1+ \beta (\Delta p)^2)$, where $\beta$ is a phenomenological parameter generalizing Newton's constant.
This leads to $\Delta x \sim \frac{1}{\Delta p} + \beta \Delta p$.  
To obtain a GUP, one should modify the position and/or momentum operators to obtain a commutator of the form
$$
[ {\hat X} , {\hat P}] = i \hbar (1 + \beta {\hat p}^2).
$$
Throughout the essay capitalized operators are modified and lower case operators are standard.
This modified commutator can be obtained from modified operators 
${\hat X} =i \hbar (1 + \beta p^2) \partial_p ~{\rm and}~ {\hat p} = p$ \cite{KMM} or ${\hat x}= i \hbar \partial _p ~{\rm and}~ {\hat P} = p\left( 1 + \frac{\beta}{3}p^2 \right)$  \cite{pedram} among other possibilities.
Using the standard connections between operator uncertainties their commutator, 
$\Delta X \Delta P \ge \left| \frac{1}{2i} \langle [{\hat X}, {\hat P}] \rangle 
\right| $ leads to the GUP of \cite{KMM}
$$
\Delta X \Delta P \geq \frac{\hbar}{2}\left| \left\langle (1+\beta (\Delta p)^2)\right\rangle\right|.
$$
However, as shown in \cite{BLS,uni-2022}, certain modified operators do not lead to a minimum length despite satisfying the same GUP. 

There are shortcomings of the above approach to incorporating a minimal length -- some aesthetic and others connected with confronting these minimal length models with experimental or astrophysical observations. On the aesthetic side there is the (too) wide freedom in choosing how the operators are modified. A simple example was mentioned above where both ${\hat p} = p ~;~ {\hat X} =i \hbar (1 + \beta p^2) \partial_p $ and ${\hat P} = p\left( 1 + \frac{\beta}{3}p^2 \right)  ~;~ {\hat x}= i \hbar \partial _p$ lead to the same commutator, but different physical consequences. To address this issue, we start from the modification of the operators first and then require that these modified operators satisfy the standard commutator. This seems to be counter the the GUP philosophy, which seems to require that the commutator must be modified, but we show this is not the case. 

Next, introducing a minimal length is usually accompanied by a violation of Lorentz symmetry and an associated modification of the standard dispersion relationship $E^2 - p^2 = m^2$. This has observable consequences as it can lead to photons dispersion -- photons of different energies having different velocities. In this essay we show that one can modified the energy and momentum operators (as well as the position and time) in such a way that one maintains the standard dispersion relationship for the modified operators. We introduce the change of having a minimal length scale, while at the same time not changing much of the standard structure of quantum mechanics and relativity as possible {\it i.e.} the position-momentum commutator and energy-momentum relationship.  \\

{\large {\bf Modified operators, standard commutator}} \\

Following the reasoning laid out above, we begin by modifying the momentum and position operators. Generally, if there is a minimal length one expects a maximum momentum/energy. Thus for our modified momentum operator we follow \cite{BJLS} and take the modified momentum in momentum space as 
\begin{equation}
    \label{p-tan}
    {\hat P}  = \frac{2 p_M}{\pi } \arctan \left( \frac{\pi p}{2 p_M} \right)~.
\end{equation}
For $p \ll p_M$, this reduces to the usual momentum operator (${\hat P} \approx p$) and for $p \gg p_M$ the momentum is capped by $p_M$. There are other modifications which satisfy the above constraints, such as ${\hat P}  =  p_M \tanh \left( \frac{p}{p_M} \right)$ \cite{BJLS}, but for brevity and concreteness we will use \eqref{p-tan} in this essay. Having chosen the modified momentum as in \eqref{p-tan}, plus the requirement that the position-momentum commutator remain the same, immediately fixes the modified position operator in momentum space to be 
\begin{equation}
    \label{x-tan}
    {\hat X} = i \hbar \left[ 1+ \left( \frac{\pi p}{2 p_M}\right)^2 \right] \partial _p ~.
\end{equation}
Substituting \eqref{p-tan} and \eqref{x-tan} into the commutator immediately gives  $[{\hat X}, {\hat P}] = i \hbar$ which implies the standard uncertainty relationship, $\Delta X \Delta P \ge \frac{\hbar}{2}$. Both \eqref{p-tan} and \eqref{x-tan}, reduce to the standard operators in momentum space in the limit when $p_M \to \infty$, namely ${\hat P} \to p$ and ${\hat X} \to i \hbar \partial _p$. Note the modified position operator in \eqref{x-tan} has the same quadratic correction term as the modified position of \cite{KMM} where ${\hat X} = i \hbar (1 + \beta p^2) \partial _p$, but with $\beta \to \frac{\pi^2}{4 p^2 _M}$.

Now the question becomes ``Do the modified operators of \eqref{p-tan} and \eqref{x-tan}, with the unmodified commutator, still give a minimum length?". Since the commutator is the same but the position and momentum are modified, the connection between the operator uncertainties and the commutator yields $\Delta X \Delta P \ge \frac{\hbar}{2}$. This seems to indicate that one would not have a minimal length. However, the modification of the operators -- and in particular the modification of the momentum -- changes the behavior of $\Delta P = \sqrt {\langle {\hat P}^2 \rangle- \langle {\hat P} \rangle ^2}$. First since $\langle {\hat P} \rangle ^2 \ge 0$ we have $\Delta P \le \sqrt {\langle {\hat P}^2 \rangle}$. Next, 
$$
\langle {\hat P}^2 \rangle = \frac{4 p_M ^2}{\pi^2} \int _{-\infty} ^{\infty} \Psi ^* (p) \arctan ^2 \left( \frac{\pi p}{2 p_M} \right) \Psi (p) dp \le p_M ^2 \int _{-\infty} ^{\infty} \Psi ^* (p) \Psi (p) dp =  p_M ^2 ~,
$$
where we have used the property that $\arctan (x) \le \frac{\pi}{2}$ and also the normalization of the momentum space wavefunction $\Psi (p)$. Putting all this together implies that the uncertainty of the modified momentum satisfies $\Delta P \le p_M$. Due to the way in which the momentum operator was modified the behavior of the uncertainty in momentum behaves differently. For the standard momentum operator, $\Delta p$ can range from 0 to $\infty$ (recall that capitalized operators are modified and lower case are standard). Combining this behavior of the modified momentum uncertainty with the unmodified uncertainty relationship gives 
$$
\Delta X \ge \frac{\hbar}{2 p_M}~,
$$
so despite the uncertainty relationship being the same, there is nevertheless a minimum $\Delta X$. However, the manner in which this minimum is obtained is different from the usual GUP case shown in figure 1. For the GUP model defined by the operators \eqref{p-tan} and \eqref{x-tan} the minimum positional resolution is arrived at asymptotically as shown in figure 2.   

\begin{figure}[H]
\label{fig2}
    \centering
    \includegraphics[scale=0.80]{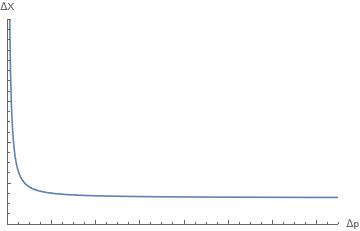}
    \caption{The relationship between $\Delta X$ and $\Delta p$ for the modified operators \eqref{p-tan} and \eqref{x-tan} where the minimum $\Delta X$ is approached asymptotically as $\Delta p \to \infty$}
\end{figure}

{\large {\bf Minimal length, photon dispersion}} \\

If spacetime has a minimum distance scale, one potential sign of this would be photon dispersion -- photons of different energies would move with different speeds through spacetime, much the same as occurs with photons traveling through a medium like water or glass. Thus seeing photon dispersion in vacuum would test for the quantum gravity motivated minimal length. Such a test was proposed in \cite{AC-nature} where it was shown that one could potentially probe the quantum gravity minimal length scale by looking at how gamma ray photons from short gamma ray bursts (GRBs) would spread out in time due to a spacetime minimum length scale. 

Introducing a minimum length scale usually alters the energy-momentum relationship for massless photons from $p^2 = E^2$ to $ p^2 = E^2 [1 + f\left(E/E_{QG}\right)]$. The function $f(E/E_{QG})$ is some energy dependent function that defines ones theory of quantum gravity, and $E_{QG}$, is the quantum gravity scale. In our modified operators from equations \eqref{p-tan} and \eqref{x-tan}, this quantum gravity scale $E_{QG}$ would be associated with $p_M$. Expanding $f(E/E_{QG})$ gives $p^2 = E^2 [1 + \xi (E/E_{QG}) + {\cal O} (E/E_{QG})^2 ]$ with $\xi$ being a parameter of order unity. The velocity of a photon is determined by $\frac{\partial E}{\partial p}$. For the standard energy-momentum relationship this gives $v= \frac{\partial E}{\partial p} =1$, while for the generically modified relationship one has $v = \frac{\partial E}{\partial p} \approx  \left(1 - \xi \frac{E}{E_{QG}} \right)$. This energy dependent photon velocity gives rise to a time difference, $\delta t$, in the arrival times of photons with different energies, $\delta E$, that is $\delta t = \xi  L \frac{\delta E}{E_{EQ}}$, with $L$ being the distance traveled by the photons from source to detector. By measuring $L$, $\delta t$ and $\delta E$ for photons coming from a distant gamma ray burst one can measure (or put limits on) $E_{QG}$.

Since the proposal in \cite{AC-nature}, there have been measurements of several short gamma ray bursts. None of these observations have detected positive evidence of photon dispersion, but these observations have been used to place limits on $E_{QG}$. The most famous of these was the 2009 gamma ray burst GRB090510 detected by the Fermi Gamma-ray Space  Telescope \cite{abdo}. This 2009 observation placed the phenomenological quantum gravity scale/the scale of the minimum distance as smaller than the Planck distance ({\it i.e.} $l_{QG} < l_{Planck}$) or larger than the Planck energy  ({\it i.e.} $E_{QG} > E_{Planck}$). This is against the expectation, that evidence of the quantum gravity scale/minimum distance scale should occur before (usually well before) one reaches the Planck scale.   

To reconcile a minimum length scale with this observational lack of photon dispersion one can require that the energy and time operators be modified in conjunction with the momentum and position operators in a way that preserves the standard energy-momentum and time-position relationship for the modified operators. In other words, following \eqref{p-tan} and \eqref{x-tan} the modified energy and time should be  
\begin{equation}
    \label{E-arctan}
    {\hat {\cal E}} = 
    \frac{2 E_M}{\pi}~\arctan{\left(\frac{\pi E}{2 E_M}\right)}~~~~;~~~~
        {\hat T } = i\hbar \left[{1+\left(\frac{\pi E}{2 E_M}\right)^2}\right] \frac{\partial}{\partial E}~.
\end{equation}
In \eqref{E-arctan} the maximum energy is given by the maximum momentum $E_M = p_M$, and since capital E is the conventional choice for standard energy, we have indicated the modified energy by a capital and script E. For the modified operators in \eqref{p-tan} and \eqref{E-arctan} the standard energy-momentum relationship for a massless photon applies, namely ${\cal E} ^2 - P^2 = 0$. Thus the photon's velocity calculated from this is $v = \frac{\partial {\cal E}}{\partial P} =1$ does not exhibit dispersion. This avoids the constraints coming from the observations of gamma ray burst photons, while still having a minimal distance. Thus one has the major change of introducing a minimal distance (as inspired by quantum gravity) while not changing the quantum commutator nor the relativistic energy-momentum relationship. \\

{\large {\bf A test for hidden change}} \\

The above construction allows one to have a minimal length (as generic arguments about quantum gravity imply) while avoiding the restrictions coming from the non-observation of photon dispersion that many theories with minimum lengths have. However this has the feel of doing theoretical contortions to avoid experimental/observational constraints. It would be more satisfying to have a positive connection between a model and experiment/observations. To this end the above model may find validation in explaining the  Greisen-Zatsepin-Kuzmin (GZK) anomaly \cite{GZK} for ultra high energy cosmic rays. 

The GZK anomaly is connected with the process whereby ultra high energy cosmic rays (generally high energy protons) interact  with photons from the CMB to produce neutral or charged pions via the interactions  $p + \gamma \to p + \pi^0$ or $p + \gamma \to n + \pi^+$. These interactions reduce the energy of the initial ultra high energy proton, making it very unlikely that one will observe such high energy cosmic rays at any great cosmological distance from their source. Simple relativistic kinematics of the process $p + \gamma \to p + \pi^0$ shows that the cosmic ray energy must satisfy the condition  
\begin{equation}
    \label{threshold}
    E_{proton} \ge \frac{(m_{proton}+ m_\pi)^2 - m_{proton}^2}{4 E_\gamma} ~,
\end{equation}
in order for the process to occur. In \eqref{threshold}, $E_{proton}$ and $E_\gamma$ are the energy of the incoming cosmic ray and CMB photons, respectively,  and $m_{proton}$ and $m_\pi$ are the rest masses of the cosmic ray proton and pion, respectively. The equality in \eqref{threshold} is the threshold value for the reaction. Putting in the rest masses and the average energy of a CMB photon one finds $E^{(threshold)}_{proton} \sim 5 \times 10 ^{19}$ eV. This is the GZK cutoff since above this threshold energy the reaction $p + \gamma \to p + \pi^0$ robs energy from the cosmic ray, and one should not see cosmic rays above this energy unless they originate ``nearby", in cosmological terms. However, cosmic rays with energies up to $3 \times 10^{20}$ eV, almost an order of magnitude above the GZK cutoff, have been observed \cite{OMG}. This is the GZK anomaly. 

Models of quantum gravity with a minimal length might hold the key to giving an answer to the GZK anomaly since these models can modify the momentum and energy which go into calculating the threshold condition in \eqref{threshold}. This idea that minimal length models which modify the energy and momentum operators might address the GZK anomaly was already considered in \cite{AC-gzk}. Simply if one had a modified energy, ${\cal E}$, modified momentum $P$ and possibly modified rest mass, $M$, which satisfied the standard relationship ${\cal E}^2 - P^2 = M^2$, then on could re-derive \eqref{threshold} in the form $ {\cal E}_{proton} \ge \frac{(M_{proton}+ M_\pi)^2 - M_{proton}^2}{4 {\cal E}_\gamma}$. Reference \cite{AC-gzk} found that if the modified ${\cal E}_{proton}$ was shifted upward relative to $E_{proton}$ that this did not help with the GZK anomaly. For example, reference \cite{MS} proposed ${\cal E} = \frac{E}{1-(E/E_{QG})}$ which has ${\cal E} > E$ which would worsen the GZK puzzle. However, for the modified energy of \eqref{E-arctan} the condition in \eqref{threshold} becomes   
\begin{equation}
    \label{threshold2}
    {\cal E}_{proton}  \approx E_{proton} - \frac{\pi ^2 E^3_{proton}}{12 E^2_M} \ge \frac{(m_{proton}+ m_\pi)^2 - m_{proton}^2}{4 E_\gamma} ~.
\end{equation}
In obtaining \eqref{threshold2} we have used the first two terms in the Taylor expansion of $\arctan$, and we have assumed that the rest masses, $m_{proton}, m_\pi$, and the photon energy, $E_{gamma}$ are so much smaller than $E_M$ that we can simply use the standard rest masses and energy. Now in this case the form of the modified energy is lower then the standard energy, ${\cal E}_{proton} < E_{proton}$, so this would address the GZK puzzle.   

In this essay we have given a path which would allow the change of having a minimal distance scale, while not changing the standard position-momentum commutator of quantum mechanics nor the standard energy-momentum relationship of relativity. The change we made was to the form of the energy-momentum and time-space operators. Keeping the standard form of the energy -momentum relationship in terms of the modified energy and momentum meant losing one experimental/observational handle on this quantum gravity inspired minimal length, namely photon dispersion. However, the minimal distance model presented in this essay does hold the possibility to address the GZK anomaly.

\end{document}